\documentclass[prl,showpacs,superscriptaddress,onecolumn]{revtex4-1}

\usepackage[pdftex]{graphicx}
\usepackage{amssymb,amsfonts,amsmath}

\begin{document}
\title{Solvent hydrodynamics speed up crystal nucleation in suspensions of hard spheres}
\author{Marc Radu}
\affiliation{Theory of Soft Condensed Matter, Universit\'e du Luxembourg, L-1511 Luxembourg, Luxembourg}
\affiliation{Institut f{\"u}r Physik, Johannes Gutenberg Universit{\"a}t, D 55099, Mainz, Germany}
\author{Tanja Schilling}
\email{tanja.schilling@uni.lu}
\affiliation{Theory of Soft Condensed Matter, Universit\'e du Luxembourg, L-1511 Luxembourg, Luxembourg}

\begin{abstract}
We present a computer simulation study on the crystal nucleation process in suspensions of hard spheres, fully taking into account the solvent hydrodynamics. If the dynamics of collodial crystallization were purely diffusive, the crystal nucleation rate densities would drop as the inverse of the solvent viscosity. However, we observe that the nucleation rate densities do not scale in this way, but are enhanced at high viscosities. This effect might explain the large discrepancy between the nuclation rate densities obtained by simulation and experiment that have reported in the literature so far.
\end{abstract}

\maketitle

\section*{Introduction}
Colloids are widely used as model systems to study fundamental questions of 
statistical mechanics. Over the past twenty-five years the phase behaviour, 
phase transition kinetics and glass transition of colloidal suspensions have 
been observed in numerous experiments and modelled by means of theory and 
simulations \cite{Loewen1994, Palberg1999, Poon2002, Anderson2002, Yethiraj2007, Gasser2009, Hunter2012}. In addition to the interest in colloids in their own right, it is often argued that colloids could serve as model systems for atomic and molecular substances \cite{Pusey1991, Poon2012}. 
Indeed, colloids can be designed to resemble atoms in many aspects of their equlibrium structure and phase behaviour. But there is a major difference in their dynamics: Colloidal particles are suspended in a solvent. They interact directly with each other (e.g.~by excluded volume) as well as indirectly by means of momentum transfer via the solvent. The latter phenomenon (``hydrodynamic interaction'') is well known and thoroughly studied in the context of colloidal flow and sedimentation \cite{Ramaswamy2001}. But when colloids are used as model systems for phase transition kinetics or for the glass transition, hydrodynamic interactions are often neglected \cite{Anderson2002, Gasser2009, Hunter2012, Loewen2009}. 

In this article we address the effect of hydrodynamic interactions on crystal nucleation in suspensions of colloidal hard spheres. Since the pioneering experiments of Pusey and van Megen in the 1980s \cite{Pusey1986} crystal nucleation in hard spheres has been observed in numerous experiments and simulations. Strikingly, the nucleation rate densities obtained by computer simulation differ by orders of magnitude from those observed in 
experiments \cite{Schaetzel1993, Harland1997, Auer2001, Filion2011, Schilling2011}. The physical reason for this discrepancy has not been understood yet.
Up to now, no simulation study on this topic has taken into 
account hydrodynamic interactions \cite{Auer2001, Filion2011, Schilling2011}.
Thus it makes sense to ask whether the solvent, which is inevitably present 
in the experiments, is the reason for the discrepancy.

Nucleation is commonly described by classical nucleation theory or extensions 
thereof. This type of theory relies on the notion of a free energy lanscape in which the system moves, i.e.~it is based on the assumption that 
during the phase transition process there is a small number of macroscopic observables which vary slowly (e.g.~the size of the largest 
crystallite), while all other degrees of freedom are equilibrated very 
quickly. Under this assumption the nucleation 
process can be modelled in terms of transition state theory: 
the nucleation rate density I(t) is then given by the product of the 
Boltzman weight of 
the height of the free energy barrier associated with the formation 
of a critical nucleus, $\Delta G^*$, and a kinetic prefactor $\kappa$. 
\[
I(t) = \kappa \exp(-\beta\Delta G^*)
\]
In the context of 
crystallization of a colloidal hard sphere suspension this implies that 
the dynamics 
of the solvent only enter the kinetic prefactor, because the height of 
the barrier 
is determinded completely by equilibrium properties of the hard spheres. 

If the process by which particles are attached to the crystal nucleus were purely diffusive, then the self-diffusion time would be the only relevant time-scale entering the kinetic pre-factor.

\section*{Results and Discussion}
We have simulated systems containing $8240$ hard spheres at volume fractions, $\phi = 0.537, 0.539$ and $0.544$ suspended in a solvent. The solvent was modelled by means of multi-particle collision-dynamics (MPCD) \cite{MalevanetsKapral1999, MalevanetsKapral2000, IhleKroll2001}. The starting configurations were prepared in the supersaturated liquid state and we verified that they did not contain crystalline precursors. Then we simulated 40 independent trajectories per value of solvent viscosity until crystallization was reached in all cases. For $\phi = 0.537$ and $0.539$ we observe an induction period that is long 
compared to the diffusion times of the system followed by a regime of 
rapid growth. Hence, for these two volume fractions, we are confident that 
we are dealing with nucleation.

When carrying out the simulations, we did not wish to make any assumptions 
on the evolution of the density of states or the length of correlation-times 
involved in the nucleation process. 
In particular, we wanted to allow for processes that might involve 
other ``slow'' coordinates than the size of the largest nucleus. Therefore -- 
in contrast 
to other simulation studies on hard spheres \cite{Auer2001, Russo2012} -- we 
did not use any free energy based sampling scheme to speed up the simulations.

We present all data in units of the sphere diameter $a$, the mass of a sphere $m$ and the thermal energy $k_BT$ (these are the intrinsic units of the MPCD program, see methods section). Solvent viscosities range between approximately $4\,\sqrt{mk_{\rm B}T}/a^2$ and $70\,\sqrt{mk_{\rm B}T}/a^2$. Translated to an experimental system with colloidal particles of radius $420\,nm$ suspended in a solvent of mass density $1\,g\,cm^{-3}$ at room temperature, these viscosities correspond to a range of $8.9\cdot 10^{-6}\,Pa\cdot s$ to $1.5\cdot10^{-4}\,Pa\cdot s$. 

\begin{figure}
  \centering
  \includegraphics[width=0.5\columnwidth]{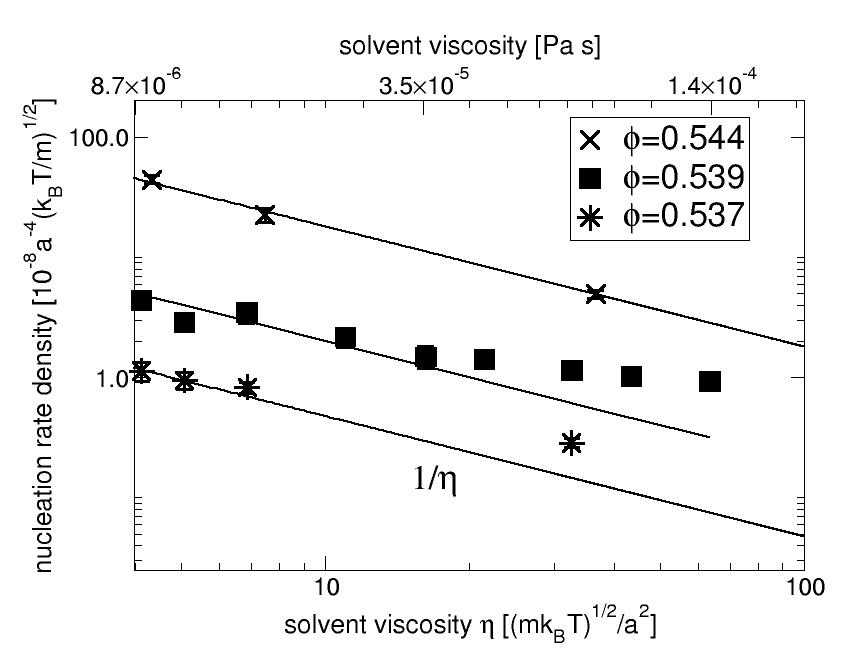}
   \caption{Nucleation rate densities as a function of solvent viscosity. If the only 
relevant time-scale were the diffusion time, the rates would drop as $1/\eta$.}
  \label{fig:IUnscaled}
\end{figure}

Fig.~\ref{fig:IUnscaled} shows the nucleation rate density versus viscosity.
If the time-scale entering the kinetic pre-factor was determined by 
the diffusion of the spheres only, the nucleation rate density would drop as 
$1/\eta$. The simulation data for the two lower volume fractions in fig~\ref{fig:IUnscaled} clearly deviate 
from a $1/\eta$-law for high viscosities. Hence we conclude that 
hydrodynamic interactions speed up the nucleation process \cite{footnote}.
  
\begin{figure}
  \centering
  \includegraphics[width=0.5\columnwidth]{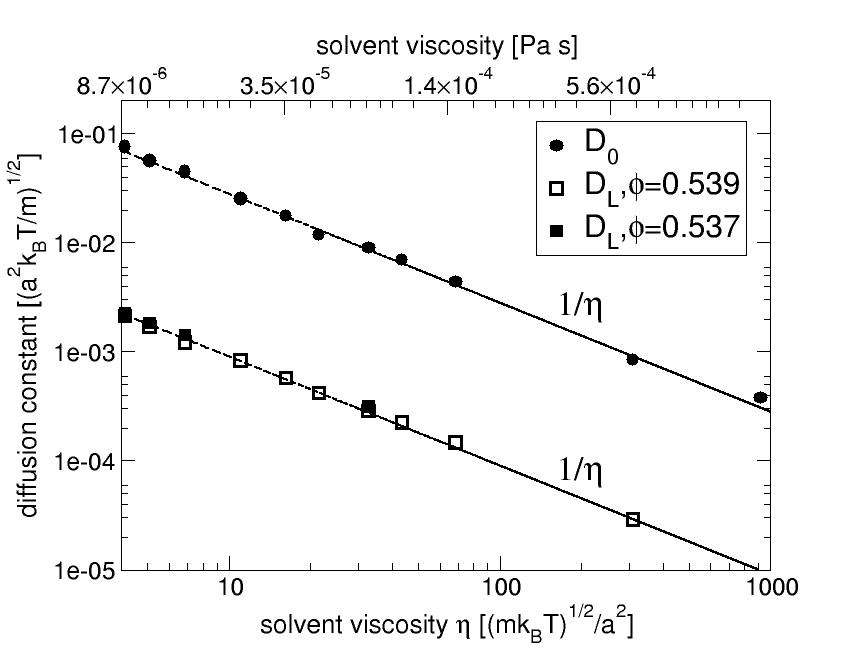}
   \caption{Diffusion constants in the infinitely dilute system and the dense suspension.}
  \label{fig:DiffCoeffs}
\end{figure}

For a consistency check  
fig.~\ref{fig:DiffCoeffs} 
shows the long-time self-diffusion constants in the infinitely dilute 
system $D_0$, and the long-time self-diffusion constant in the 
supersaturated suspension $D_L$. Both diffusion constants follow the 
expected $1/\eta$-behaviour.

Experimentally, hard sphere suspensions are synthesized in various ways. 
Common systems are polystyrene spheres suspended in water, and sterically 
stabilized polymethylmethacrylate (PMMA) spheres in an organic liquid such 
as decalin. Typical solvent viscosities are in the range of $1\cdot 10^{-3} - 
3\cdot 10^{-3}\;Pa\;s$. When simulation data is compared with experiments -- 
or when experiments of chemically different composition 
are compared to one another -- the solvent is 
usually taken into account by normalizing the nucleation rate density with respect 
to either $D_L$ or $D_0$ \cite{Auer2001,Filion2011,Schilling2011}. As 
argued above, this 
procedure is based on the assumption that the nucleation process can be 
described by transition state theory with a kinetic prefactor that contains
purely diffusive attachment dynamics -- fig.~\ref{fig:IUnscaled} clearly 
shows that this assumption does not hold. 

\begin{figure}
  \centering
  \includegraphics[width=0.5\columnwidth]{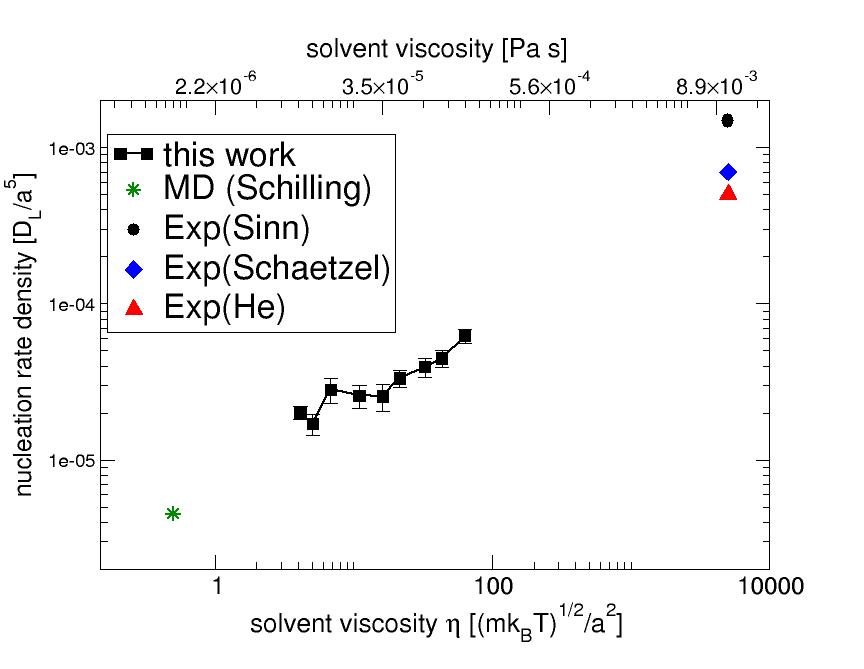}
   \caption{Nucleation rate densities scaled by $D_L$, the long-time self 
diffusion constant ($\phi=0.539$). The symbols indicate a simulation 
result without a solvent, i.e.~$\eta=0$ \cite{Schilling2011} (star) and experimental results \cite{Schaetzel1993, Sinn2001, He1996}.}
  \label{fig:ID0DL}
\end{figure}

For a quantitative comparison between simulation and experiment, 
fig.~\ref{fig:ID0DL} shows the nucleation rate densities rescaled with respect to 
$D_{\rm L}$. The green star is a computer simulation result for hard spheres 
without a solvent, in which the spheres moved ballistically and 
collided elastically. (The star hence corresponds to $\eta=0$. We placed it 
at a very small, 
non-zero value of $\eta$ instead to include it in the log-log presentation). 
The other symbols indicate experimentally measured nucleation rate densities. We 
conclude that the long debated difference 
between simulation results and experimental results is most probably due to 
solvent hydrodynamics.

To validate this statement it would be desirable to compare 
nucleation rate densities for different values of supersaturation. The discrepancy 
between the experimental data and the simulation results increases for 
decreasing 
supersaturation\cite{Auer2001,Schilling2011}. We could simulate 
lower supersaturations because the computation times became forbiddingly long, but we have computed nucleation rate densities for a higher supersaturation (see crosses in fig~\ref{fig:IUnscaled}.)  For $\phi=0.544$ the data follows a $1/\eta$-law. 
Hence, in agreement with the data shown in refs.~\cite{Auer2001,Schilling2011} 
the effect vanishes at very high supersaturations.        

\begin{figure}
  \centering
  \includegraphics[width=0.5\columnwidth]{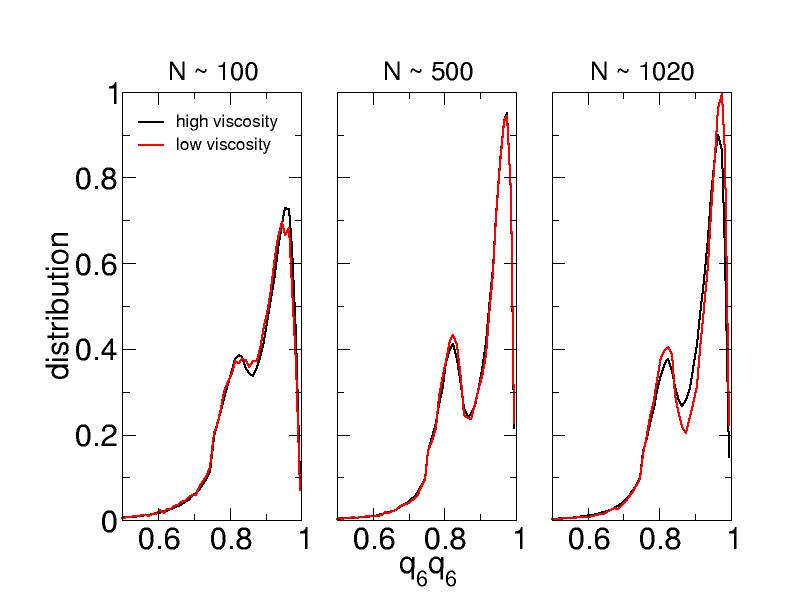}
   \caption{Distributions of $q_6q_6$ for different sizes of the largest cluster. Here shown for small and high viscosity.}
  \label{fig:Q6Q6Distr}
\end{figure}

As the nucleation rate densities are affected by hydrodynamic interactions, one could 
expect to observe differences in the sizes, shapes or structures of the 
crystallites that 
form, too. We analysed the structures of the growing crystallites in terms of 
their $q_6q_6$-bond-order \cite{Steinhardt1983, Wolde1995}. Fig.~\ref{fig:Q6Q6Distr} shows $\vec{q_6}(i)\cdot\vec{q_6}(j)$ for pairs of particles $i$ and $j$ in clusters of equal sizes obtained at different solvent viscosities. The crystallites are very similar in structure. Within the statistical accuracy, the radii of gyration of the crystallites did not differ, either. We also performed a  
committor analysis for the highest and the lowest viscosity and did not 
find any difference in the critical cluster 
size, shape or structure. This observation proves that a description 
of the crystal nucleation process in terms of transition state theory is 
adequate. Hydrodynamic interactions, however, influence the kinetic pre-factor.

\section*{Conclusion}
In summary, we have simulated crystallization from a supersaturated liquid 
suspension of hard spheres taking into account the solvent hydrodynamics. 
We find that kinetics need to be taken with care when one 
studies phase transitions in colloids. Contrary to what has been assumed in 
the literature so far, the crystal nucleation rate densities do not drop as 
the inverse viscosity, i.e.~the attachment dynamics are not purely diffusive. 
The structure and shape of the critical nucleus are independent of solvent 
viscosity, 
hence a description in terms of transition state theory is adequate. 
But the kinetic pre-factor is affected by hydrodynamic interactions. 
It would be very interesting to see a test of this effect in an experiment on hard spheres suspended in a solvent of a different chemical composition than commonly used.

\section*{Simulation methods and Analysis}
To simulate hard spheres suspeded in a liquid, we used a combination of an event-driven molecular dynamics (EDMD) algorithm \cite{Alder1959,MarinCordero1995,Krantz1996,Lubachevsky1991} for the spheres and multiparticle-collision dynamics (MPCD)  \cite{MalevanetsKapral1999,MalevanetsKapral2000} as a mesoscopic solvent model to account for the hydrodynamic interactions.
The basic idea of a MPCD algorithm is to transport momentum through the system by means of point particles of mass $m$ while satisfying the conservation laws of mass, energy and momentum locally. 
 The algorithm consists of two steps, namely free streaming interrupted by multiparticle collisions. In the streaming step, all fluid particles are propagated ballistically for a time-interval of length $h$. Then the fluid particles are sorted into a lattice of cubic cells of size $a\times a\times a$ and the particle velocities are rotated around the center of mass velocity in this cell. Before a collision step is carried out, the collision cell grid is shifted by a randomly chosen vector with components taken from the interval $\left[-a/2,a/2\right]$, to ensure Galilean invariance \cite{IhleKroll2001}. The solvent viscosity is varied by varying the duration $h$ of the ballistic flight.

In order to measure the solvent viscosities we imposed a Poiseuille flow between two planar walls. From the resulting parabolic velocity field we extracted $\eta$.

\subsection*{Analysis}

\underline{Identification of crystallites}\\
We identified crystallites by means of the ``q6q6-bond order parameter'' \cite{Steinhardt1983, Wolde1995}. 
For a sphere $i$ with $n(i)$ neighbors, local 6-fold bond-orientational order 
is characterized by
\[
\bar{q}_{6m}(i) := \frac{1}{n(i)}\sum_{j=1}^{n(i)} Y_{6m}\left(\vec{r}_{ij}\right)\quad ,
\]
where $ Y_{6m}\left(\vec{r}_{ij}\right)$ are the spherical harmonics for $m=-6 \dots 6$ and $\vec{r}_{ij}$ is the position vector between 
a sphere $i$ and its neighbour $j$.
A vector $\vec{q}_{6}(i)$ is assigned to each sphere, the elements 
 of which are defined as 
\begin{equation}
q_{6m}(i) := \frac{\bar{q}_{6m}(i)}{\left(\sum_{m=-6}^6|\bar{q}_{6m}(i)|\right)^{1/2}} \quad .\label{BOP}
\end{equation}
If a sphere had more than 9 neighbours with  $\vec{q_6}(i)\cdot \vec{q_6}^*(j) > 0.7$ it was considered ``crystalline''.\\
\\
\underline{Nucleation times}\\
Once a run had produced a crystalline cluster of more than 80 particles, it 
definitely crystallized. 
Thus we used this value to locate the nucleation time. To test the validity 
of this 
criterion, we performed a commitor analysis for two values of viscosity 
($\eta = 4.17 \sqrt{mk_BT}/a^2$ and $\eta = 63.93 \sqrt{mk_BT}/a^2$). We 
found that a cluster size of ca.~30 particles corresponds to a 50\% 
probability for subsequent full crystallization in both cases. As the growth process is very fast, the induction times extracted from the commitor analysis hardly differ from those obtained by means of the ``80-particle criterion''. 
We took the arithmetic mean of the distribution of measured induction times $\langle t_{\rm ind} \rangle$ to determine the nucleation rate density, and its standard deviation to 
determine the error bars. The nucleation 
rate density is then given by 
\[ I = \frac{1}{V \langle t_{\rm ind}\rangle} \quad ,\]
where V is the volume of the system.\\
\\
\underline{Diffusion constants}\\
To determine $D_0$ ($D_L$) we computed the mean squared displacement of a 
particle in the infinitely dilute (resp.~the dense) solution versus time and 
fitted a straight line to it. The result for $D_0$ was consistent with the 
shear 
viscosity  $\eta$ that we had measured independently, see above. I.e. the 
Stokes-Einstein relation was fullfilled. 

\section*{Acknowledgments}
We thank William van Megen, Peter Pusey, Hajime Tanaka, Hajo Sch\"ope and Tobias Kraus for providing us with data on solvent viscosities, and Paul van der Schoot for helpful comments on our results. This work has been supported financially by the German research foundation (DFG)  within SFB TR6. The present project is supported by the National Research Fund,
Luxembourg.




\end{document}